\documentstyle[epsf,eqsecnum,floats,preprint,aps,epsfig]{revtex}

\input epsf
\tighten
\begin{document}
\newcommand{\bm}[1]{\mbox{\boldmath{$#1$}}}
\newcommand{\be}{\begin{equation}}
\newcommand{\ee}{\end{equation}}
\newcommand{\bea}{\begin{eqnarray}}
\newcommand{\eea}{\end{eqnarray}}
\newcommand{\barr}{\begin{array}}
\newcommand{\earr}{\end{array}}

\newcommand{\barz}{\bar z}
\newcommand{\tillam}{\tilde \lambda}
\newcommand{\tlam}{{\tau_\lambda}}
\newcommand{\sinlam}{\sin \tilde \lambda \pi}
\newcommand{\co}{{\cal O}}

\newcommand{\quar}{\frac{1}{4}}
\newcommand{\half}{\frac{1}{2}}

\rightline{UFIFT-HEP-03-29}
\rightline{hep-th/0311179}
\vskip 1cm

\begin{center}
\ \\
\large{{\bf One-loop Evolution of a Rolling Tachyon}} 
\ \\
\ \\
\ \\
\normalsize{ Xingang Chen\footnote{\tt email address: xgchen@phys.ufl.edu} }
\ \\
\ \\
\small{\em Institute for Fundamental Theory \\ Department of Physics,
University of Florida, Gainesville, FL 32611 }

\end{center}

\begin{abstract}
We study the time evolution of the one-loop diagram in Sen's rolling
tachyon background. We find that at least in the long cylinder case
they grow rapidly at late time, due to the exponential
growth of 
the timelike oscillator terms in the boundary state. This can also be
interpreted as the virtual open string pair creation in the decaying
brane. This behavior indicates a breakdown of this rolling tachyon
solution at some point during the evolution. We also discuss the
closed string emission from this one-loop diagram, and the evolution
of a one-loop diagram connecting a decaying brane to a stable brane,
which is responsible for the physical open string creation on the
stable brane.
\end{abstract}

\setcounter{page}{0}
\thispagestyle{empty}
\maketitle

\eject

\vfill

\baselineskip=18pt

\section{introduction}
Brane decay \cite{Sen:1999mg} is an important process in string
theory. But the 
dynamics of this process is
difficult to study at a fundamental level because it is both
non-perturbative and of high energy. 
Recently Sen \cite{Sen:2002nu,Sen:2002in} has proposed a boundary
conformal field 
theory (BCFT) description of the rolling tachyon which opens up a
way to study the string dynamics in brane decay more quantitatively.

Sen proposed that adding a timelike sine-Gordon type boundary operator
at open string end corresponds to a tachyon rolling down from an
inverted 
potential. The corresponding boundary state is exactly solvable and
has a marginal parameter describing the starting place of the rolling
tachyon. Using this boundary state, he showed that, in the absence of
the closed string coupling, the unstable brane decays into a
pressureless matter called tachyon
matter\cite{Sen:2002nu,Sen:2002in,Sen:2002an}.

To better understand the nature of this tachyon
matter, closed string coupling is studied in
\cite{Okuda:2002yd,Chen:2002fp,Lambert:2003zr}. By calculating the
one-point function on a disk in the rolling tachyon background, it is
found\cite{Lambert:2003zr,Gaiotto:2003rm} that a coherent
state of heavy closed strings is produced during the brane decay. For
the D$p$-brane with $p\le 2$, the leading order of the emitted energy
is infinite. Therefore all the energy goes to 
the closed strings localized at the original place of the unstable
brane.\footnote{More precisely, if we cut off the closed
string energy at the order of $1/g$ so that it does not exceed the
brane tension,
for D0-brane, the emitted energy is of the same order of the
brane. For $p>0$ the emitted energy after this cutoff is of lower
order in $g$.} 
This is the closed string
description of the tachyon matter\cite{Sen:2003bc,Sen:2003iv}. For $p
>2$, the emitted energy is
finite and thus of lower order in string coupling comparing to the
brane tension. It is argued \cite{Lambert:2003zr,Gaiotto:2003rm} that
physically since the long wavelength tachyon
modes will grow and eventually make these branes become causally
disconnected D0-branes\cite{Sen:2002vv,Larsen:2002wc}, they will also
all decay into closed
strings. But how fast this can happen presumably depends on the
initial
homogeneity condition of the brane, and if we start with sufficiently
homogeneous brane, the tachyon matter may still show up.

The mini-superspace
approximation\cite{Braaten:qs,Braaten:np,Polchinski:1990mh}
has been
used\cite{Strominger:2002pc,Gutperle:2003xf,Maloney:2003ck,Fredenhagen:2003ut}
to study the open string creation from the rolling 
tachyon. In this approximation the size of the open string is
neglected and the tachyon vertex operator corresponds
to an exponentially growing potential for the open string. It is found
that the production of the open strings from this background grows 
exponentially for large time $t$. The
closed string coupling is not
considered in this approximation. However since the end product of the
decaying
brane no longer have the open string degrees of freedom, the closed
strings have to be produced from the open strings. 

In this paper we
consider the virtual open string pair production and its subsequent
coupling to closed strings. This is the closed string production from
a cylinder diagram with both ends on the decaying
brane.\footnote{Another possible way of getting closed strings from
open strings is discussed in \cite{Yi:1999hd,Bergman:2000xf}.} This
diagram is the quantum 
correction, at the order of the string coupling $g$, of the one-point
function on the
disk.\footnote{Loop corrections and multiple-point functions for
decaying brane in two dimensional string theory is
recently studied in \cite{Gutperle:2003ij}. Another related discussion
on the loop diagram is in \cite{Okuyama:2003jk}.}

We first study the evolution of this cylinder diagram. We restrict
ourself to the long cylinder case where the analytical results are
simple. As we will see, this diagram grows rapidly for large $t$, due
to the exponentially growing timelike oscillator
modes\cite{Okuda:2002yd} in Sen's boundary
state. This is the indication that, if we consider the closed string
emission from this quantum effect, the brane energy will all be
converted to closed strings, regardless of the spatial dimension of
the brane. This also indicates that the back reaction will have to
modify
the late time behavior of this boundary state, especially these
timelike 
oscillator modes, i.e.~these growing modes have to die down when all
the energy is emitted.

By simply imposing a cutoff on the time evolution of the boundary
state, 
we estimate the magnitude of the closed string production. We find it
diverges very rapidly as the string level increases and may overwhelm
the classical results. 
However the details are not
clear yet, since we only have results for long cylinder case and the
cutoff is too simple. 

Open string creation will become physical if a
stable brane is added. We will study the evolution
of a one-loop diagram with one end on the rolling tachyon state and
another on the Neumann boundary state. This diagram is
responsible for the open string production on the stable brane
from the decaying brane. 

This paper is organized as follows. In Sec.~\ref{SectRevSen} we review
Sen's construction of the rolling tachyon boundary state. In
Sec.~\ref{SectSpace} we illustrate our calculation for a long cylinder
in a space-like case with both ends on a D-brane. We use this method to
study the one-loop evolution in rolling tachyon in
Sec.~\ref{SectEvolution}, we also discuss the closed
string production
and the one-loop evolution if one end is replaced by the Neumann
boundary condition. Sec.~\ref{SectCon} is the conclusion. In Appendix
\ref{AppOnePoint}, we calculate the one-point function on a cylinder
exactly with various simple boundary conditions and compare them with
the result in Sec.~\ref{SectSpace}. In Appendix \ref{AppClose}, we
supply some calculations skipped in Sec.~\ref{SectClosed}. 
For simplicity we
restrict our discussions to the bosonic string theory.

\section{Review of Sen's description}
\label{SectRevSen}
We first review Sen's BCFT description of the rolling
tachyon. We start by introducing the exact
solution\cite{Callan:1994ub,Polchinski:my} of a spacelike CFT with
a sine-Gordon boundary interaction at $\sigma=0$. The action
of this BCFT is given by 
\be
S = \frac{1}{2\pi} \int dz d\barz ~\partial X {\bar \partial} X 
    - \tillam \int d\tau \cos X(\sigma=0) ~.
\label{XBoundAct}
\ee
The exact solution of the boundary state is found to be
\be
| B \rangle = \sum_j \sum_{m=-j}^j ~ {\cal D}_{m,-m}^j
|j;m,m \rangle\rangle ~,
\label{XBoundState}
\ee
where $|j;m,m \rangle\rangle$ is the Ishibashi
states\cite{Ishibashi:1988kg} associated with the
primary states $|j;m,m \rangle = |j,m\rangle_L |j,m\rangle_R$, and
${\cal D}_{m,-m}^j$ is the SU(2)
rotation matrix element
\be
{\cal D}_{m,-m}^j = \langle j,m| e^{i\pi \tillam(J_+ +J_-)} |j,-m
\rangle ~.
\ee
It is useful to express the states
$|j,m\rangle_L \sim (J_-)^{j-m} |j,j\rangle_L$ explicitly in terms of the
oscillators and zero modes using 
the chiral SU(2)
generators\cite{Witten:1991zd,Klebanov:1991hx,Ohta:1992zp}
\bea
J_{\pm} = \oint \frac{dz}{2\pi i}~ e^{\pm 2i X(z)} ~, \qquad
J_3 = \oint \frac{dz}{2\pi i}~ i\partial X(z) ~,
\label{SU2gen}
\eea
and
\bea
|j,j \rangle_L = e^{2 i j X(0)} |0\rangle ~.
\label{Statejj}
\eea
Similar expressions hold for the anti-holomorphic states
$|j,m\rangle_R$.

The physically inequivalent solution is described by $-\half \le
\tillam < \half$. The $\tillam=0$ case
corresponds to the Neumann boundary condition. For $\tillam = -\half$,
this boundary state describes a periodic array of D-branes located at
$x=2n\pi$. For $\tillam = \half$, this D-brane array is shifted to
$x=(2n+1)\pi$. General $\tillam$ interpolates between the Neumann and
Dirichlet boundary conditions. 

Each primary state can be expanded into a zero mode $e^{2imx}$ times
oscillators of 
level $j^2-m^2$ using (\ref{SU2gen}) and (\ref{Statejj}). For example,
the oscillator-free term in (\ref{XBoundState}) is contributed by
the primary states $|j;\pm j,\pm j \rangle$. It is given by
\bea
\tilde f(x) &=& 1 + \sum_{j=\half, 1, \dots} (-\sin\tillam \pi)^{2j} 
\left( e^{2ijx} + e^{-2ijx} \right)   \label{xSeries} \\
&=& \frac{1}{ 1+ e^{ix} \sin(\tillam \pi)} + \frac{1}{ 1+ e^{-ix}
\sin(\tillam \pi)} - 1 ~,
\label{xSum}
\eea
where we denote $x$ as the zero mode of $X$.

After the inverse Wick rotation $X
\rightarrow -iX^0$,
the boundary interaction term in (\ref{XBoundAct}) becomes a
tachyon vertex operator 
\be
T(X^0) = \tillam \cosh X^0 ~.
\ee
In the vicinity of $t=0$, this corresponds to a tachyon field $T(t)$
with 
\be
T(t=0) = \tillam ~, \qquad \dot T (t=0) = 0  ~,
\ee
where $t$ is the zero mode of $X^0$. Since
this time-like BCFT is also 
exactly solvable by inverse Wick 
rotating (\ref{XBoundState}), the corresponding boundary state
together with the spatial and ghost parts
\bea
|B\rangle = {\cal N}_p~ |B\rangle_{X^0} \otimes |B\rangle_{\vec X}
\otimes |B\rangle_{bc}
\eea
should
describe the long time evolution of the rolling tachyon.\footnote{In
this paper, we consider the full-brane, which is time reflection
symmetric about $t=0$. The $t<0$ part describes the formation of a
unstable brane, and the $t>0$ part describes the decay of the
brane. The boundary state of the half-brane case is given
in\cite{Strominger:2002pc,Larsen:2002wc,Gutperle:2003xf}. }

We notice that, after the inverse Wick rotation, the series
(\ref{xSeries}) is only convergent for $t<|\log\sin{\tillam
\pi}|$. However since this series can be summed into a closed form
(\ref{xSum}), the inverse Wick rotation is well defined for all $t$ in
(\ref{xSum}). This term together with the next higher level term
\bea
|B\rangle_{X^0} = f(t)|0\rangle + g(t) \alpha^0_{-1} \tilde\alpha^0_{-1}
|0\rangle + \dots ~,
\label{tBoundState}
\eea
where
\bea
f(t) &=& \frac{1}{ 1+ e^{t} \sin(\tillam \pi)} + \frac{1}{ 1+ e^{-t}
\sin(\tillam \pi)} - 1 ~,
\label{Formft} \\
g(t) &=& \cos(2\tillam\pi) + 1 -f(t) ~,
\eea
are especially interesting because they are related to the
stress tensor of the decaying brane by
\bea
T_{00} = \half T_p (f(t) + g(t))~, \qquad T_{ij} =
-T_p f(t)\delta_{ij}~, 
\eea
where $T_p$ is the brane tension and $i,j$ denotes the longitudinal
direction of the 
brane. Therefore, in the absence of the string coupling $g$, this
boundary state
describes a system with conserved energy and becoming pressureless as
$t\rightarrow \infty$. The special case $\tillam=\half$ corresponds
to delta functions on the
imaginary axis and has no support on real $t$-axis. Correspondingly
both $f(t)$ and $g(t)$ vanish.

If we consider higher oscillator level terms in (\ref{tBoundState}),
for a given level $l$, there are only finite number of primary states
satisfying $j^2-m^2<l~(m\neq \pm j)$. Therefore the sum of them is still
exponentially dependent of $t$\cite{Okuda:2002yd}, unlike the $f(t)$
in (\ref{Formft}) whose $t$-dependence is much milder.

Since the timelike oscillator part of a on-shell closed string state
should be 
gauged away\cite{Lambert:2003zr,Hwang:1991an,Evans:1998qu}, the above
mentioned exponentially growing modes do not show up in the
one-point function of the disk amplitude for physical string
emission. However it can contribute
when an off-shell closed string is produced and affect the on-shell
closed string production from a one-loop diagram. This will be the
focus of this paper. One of the difficulties is that this diagram will
sum over a series of exponential terms involving powers of
$e^t$. If this summation diverges for a given $t$, one cannot conclude
anything since a redefinition of the summation may be needed to bring
it to a closed form like we did for (\ref{Formft}). The main
observation in this paper is that at least for the long cylinder
case we find the summation to be convergent, and that the
time-dependent behavior for this case is very different from that in
the disk case.

\section{Spacelike One-loop}
\label{SectSpace}
As we have seen in Sec.~\ref{SectRevSen}, the boundary state
(\ref{XBoundState}) defined 
by the BCFT (\ref{XBoundAct}) has a spatial dependence. For the
oscillator-free component, the Fourier modes of such a
dependence are given by the one-point function $\langle e^{ikX}
\rangle$ on the disk. In this section to illustrate our method we will
study the amplitude $\langle e^{ikX} \rangle$ on an Euclidean
cylinder.

Since the boundary state is known, this open string one-loop
diagram can be calculated in terms of closed strings. A state
corresponding to the vertex operator $e^{ikX}$ is emitted during the
closed 
string propagation from one boundary state to another. The amplitude
is then
\bea
\tilde A_{BB}(k) &=& \langle B| q_1^{L_0 + \tilde L_0} e^{ikX}
q_2^{L_0 + \tilde L_0} |B\rangle  \nonumber \\
&\equiv& \int dx ~e^{ikx}~ A_{BB}(x) ~, 
\label{OneloopDef}
\eea
where $q_{1,2} = e^{-2\pi s_{1,2}}$ and the $x$ is the zero-mode of
the $X$. In the second line we have denoted the Fourier transform of
this amplitude as $A_{BB}(x)$. $A_{BB}(x)$ is also dependent of the
moduli $s_{1,2}$. 

For a boundary state (\ref{XBoundState}) with general $\tilde
\lambda$, this is still difficult to calculate since the vertex
operator $e^{ikX}$ includes all the oscillator modes which must be
contracted with those from the boundary states.
But the calculation is simplified as $s_{1,2}$ become large. Comparing
to the zero mode contribution in $e^{ikX}$, the oscillator
contributions are  of higher order in $q_{1,2}$, thus get suppressed if
$s_{1,2} \gg 1$. That is, we will study the following amplitude
\bea
\langle B| q_1^{L_0 + \tilde L_0} e^{ikx}
q_2^{L_0 + \tilde L_0} |B\rangle ~.
\label{OneloopApprox}
\eea

To give an example we study the case of a single D0-brane located at
$x=0$
\bea
|D\rangle = \delta(x) ~ e^{\sum_{n=0}^\infty \frac{1}{n}
\alpha_{-n} \tilde \alpha_{-n} } |0\rangle ~.
\eea
Using the approximation (\ref{OneloopApprox}) we get
\bea
\tilde A_{DD}(k) &\approx& \langle D| q_1^{L_0 + \tilde L_0} e^{ikx} 
q_2^{L_0 + \tilde L_0} |D\rangle   \nonumber \\
&\approx& \frac{1}{4\pi^2}  \int dp~ q_1^{(p-k)^2/2}
q_2^{p^2/2}
\label{ADDk1} \\
&=& \frac{1}{4\pi^2}  \frac{1}{\sqrt{s}}
e^{-\pi k^2 s_1 s_2/s}~. \qquad (s=s_1+s_2)
\label{ADDk}
\eea
Fourier transforming it to the coordinate space, we get
\bea
A_{DD}(x) &\approx& \frac{1}{8\pi^3}
\int dp_1 dp_2~ q_1^{p_1^2/2} q_2^{p_2^2/2} e^{i(p_1-p_2)x} 
\nonumber \\
&=& \frac{1}{8\pi^3} \frac{1}{\sqrt{s_1 s_2}}
e^{-\frac{x^2}{4\pi} \left( \frac{1}{s_1} + \frac{1}{s_2} \right)} ~.
\label{ADDx}
\eea
Different from the disk amplitude, the one-loop amplitude is no longer
localized at $x=0$. Physically this
is due to the closed string propagating off the D-brane and
then coming back. This propagation is parameterized by the moduli
$s_{1,2}$. From (\ref{ADDx}) we can see that as $s_{1,2}$
decrease the amplitude gets more localized toward the location of the
brane.

In Appendix \ref{AppOnePoint} we will verify (\ref{ADDk}) using the
exact result.

\section{One-loop evolution}
\label{SectEvolution}

\subsection{One-loop for rolling tachyon}
\label{SectRolling}
To study the closed string production from the cylinder diagram, we
will be interested in the quantity $\langle e^{iEX^0} \rangle_{\rm
cylinder}$. The boundaries of this cylinder are given by Sen's rolling
tachyon boundary state $|B\rangle_{X^0}$. This quantity gives
the timelike component of the physical closed string emission. Similar
to Sec.~\ref{SectSpace}, we will denote the Fourier transform of
$\langle e^{iEX^0} \rangle_{\rm cylinder}$ as $A_{BB}(t)$ and think of
it as a time evolution of the one-loop diagram with moduli
$s_{1,2}$. Again, we will concentrate on the long
cylinder case where we only take the zero mode of this vertex
operator,\footnote{An exception is discussed in footnote
\ref{FTexception}.} namely $e^{iEt}$. We will
be interested in the small $\tillam$ case and this corresponds to
placing the tachyon near the top of the potential.

Using the explicit form of the boundary state $|B\rangle_{X^0}$ given
by inverse Wick rotating (\ref{XBoundState}), we have
\bea
\tilde A_{BB}(E)
&\equiv& \int dt ~e^{iEt} A_{BB}(t)  
\label{ABBEa} \\
&\approx&_{X^0}\langle B| 
q_1^{L_0 + \tilde L_0}~e^{iEt}~q_2^{L_0 + \tilde L_0}
|B\rangle_{X^0} 
\label{ABBEb} \\
&=& \sum
{\cal D}^{j_1~\dagger}_{m_1,-m_1} ~ {\cal D}^{j_2~}_{m_2,-m_2} 
~{}_{X^0}\langle\langle j_1;m_1,m_1|~  
q_1^{L_0 + \tilde L_0}~e^{iEt}~q_2^{L_0 + \tilde L_0}
~|j_2;m_2,m_2 \rangle\rangle_{X^0} ~.  
\label{ABBE}
\eea
Note an integration over $t$ is implicit in the last two
lines.
Since only the zero mode $e^{iEt}$ enters in (\ref{ABBE}), $A_{BB}(t)$
can be read off from (\ref{ABBE}) before doing the integration $\int
dt~e^{iEt} \cdots$. However as we will see, although $A_{BB}(t)$ turns
out to be convergent for a given $t$, it blows up as $t\rightarrow
\infty$. Therefore a upper limit of the integration over $t$ has to be
specified on physical grounds.

To perform the summation in (\ref{ABBE}),
we have to distinguish two different cases.
The first case is when $m_1=\pm j_1$ and
$m_2=\pm j_2$. We denote this subspace of $|B\rangle_{X^0}$ as
$|B^0\rangle$. As we see 
from Sec.~\ref{SectRevSen}, for the primary
states, this is an
infinite series of oscillator-free terms. It only makes sense if we
sum them over to a closed form before we do the inverse Wick
rotation. Therefore in
(\ref{ABBE}) we should use the form (\ref{Formft}) rather than treat
these terms individually. The descendent
states can be built by acting the Virasoro generators on this primary
state
\bea
\sum_{\{n_m\}} \prod_{m>0} 
\left( L_{-m} \tilde L_{-m} \right)^{n_m} f(t)
|0;0,0\rangle ~.
\label{Case1Sum}
\eea
In our long cylinder case, to get the leading term we only need to
look at the primary states, because the descendent states give rise to
terms higher order in $q_{1,2}$.
Hence We get
\bea
\langle B^0| 
q_1^{L_0 + \tilde L_0}~e^{iEt}~q_2^{L_0 + \tilde L_0}
|B^0 \rangle 
\approx  \int dt~ e^{iEt} 
\left( q_1^{\half \frac{\partial^2}{\partial t^2}} f(t) \right) 
\left( q_2^{\half \frac{\partial^2}{\partial t^2}} f(t) \right)~.
\label{ABBE1a}
\eea
So this case contributes to $A_{BB}(t)$ a term
\bea
A^{B^0}_{BB}(t) \approx
\left( q_1^{\half \frac{\partial^2}{\partial t^2}} f(t) \right) 
\left( q_2^{\half \frac{\partial^2}{\partial t^2}} f(t) \right)~.
\label{ABBE1}
\eea

The second case includes
the rest of the terms in $|B\rangle_{X^0}$. We denote it as $|\tilde B
\rangle$. The contraction of these
terms give\footnote{There are no cross terms between $|B^0\rangle$ and
$|\tilde B\rangle$ in (\ref{ABBEb}) since the corresponding primary
states have different oscillator levels.}
\bea
&&\langle \tilde B|
q_1^{L_0 + \tilde L_0}~e^{iEt}~q_2^{L_0 + \tilde L_0}  
|\tilde B \rangle
\nonumber \\ 
&\approx& \sum_{m_{1,2} \neq \pm j_{1,2}}
{\cal D}^{j_1~\dagger}_{m_1,-m_1} ~ {\cal D}^{j_2~}_{m_2,-m_2}
~\langle j_1;m_1,m_1 | e^{iEt} | j_2;m_2,m_2 \rangle ~ 
q_1^{2j_1^2} q_2^{2j_2^2} ~.
\label{ABBE2a}
\eea
The factor $q_1^{2j_1^2} q_2^{2j_2^2}$
is obtained by acting $q_{1,2}^{L_0 + \tilde L_0}$ on the
primary states.\footnote{Note that here we have kept the different
powers of $q_{1,2}$, each of which is the leading term, contributing
by the primary states, over those from the descendent states
for small $q_{1,2}$. However among these terms in (\ref{ABBE2a}), it
is no longer true that the lowest $j_{1,2}$ gives the leading
behavior. This is because, as we will see shortly, the quantity
$\langle j_1;m_1,m_1 | e^{iEt} | j_2;m_2,m_2 \rangle$ exponentially
grows as $j_{1,2}$ grows. Also because of this, in (\ref{ABBE2a}) and
the subsequent (\ref{ABBE2}), we will be interested only in the
leading term.}
For small $\tilde \lambda$, 
\bea
{\cal D}^j_{m,-m}\approx
\frac{(j+|m|)!}{(j-|m|)!(2|m|)!} (i\sinlam)^{2|m|} ~.
\eea 
The primary states $|j;m,m \rangle$ consist of the zero modes
$e^{2mt}$
and the oscillator modes of level $j^2-m^2$. So in (\ref{ABBE2a}) the
quantity 
\bea
&&\langle j_1;m_1,m_1 | e^{iEt} | j_2;m_2,m_2 \rangle
\nonumber \\
&=& n(j_1,j_2,m_1,m_2)~ \delta_{j_1^2-m_1^2,~j_2^2-m_2^2}
\int dt ~e^{iEt}
~e^{2(m_2 + m_1)t},
\label{PrimContr}
\eea
where the oscillator levels have to match to give a non-zero
answer\footnote{\label{FTexception}Here the vanishing contribution of
the $j_1^2-m_1^2 \neq j_2^2-m_2^2$ case (including the cross term
between $|B^0\rangle$ and $|{\tilde B} \rangle$) leaves the
possibility of the contribution from the oscillator terms in
$e^{iEX^0}$ or the descendent states (they are needed to match the
oscillator level for $j_1^2-m_1^2 \neq j_2^2-m_2^2$). Following the
same spirit, one can see that their leading contribution comes from
terms with minimal oscillators for fixed $e^{2m_{1,2}t}$, namely
$j_1=m_1+1, j_2=m_2+1$ ($j_1 \neq j_2,~m_{1,2}>0$). After summation,
its behavior
($\sim e^{(t-\tau_\lambda)^2 (1/s_1+1/s_2)/4\pi}$) is similar to
(\ref{ABBtAppr}).}
and we use $n(j_1,j_2,m_1,m_2)$ to denote the number resulted from the
contraction between two normalized oscillator states of the same
level.
Besides some special values of $j$ and $m$, the delta function
in (\ref{PrimContr}) is non-zero for $j_1=j_2=j$ and $m_1=\pm m_2$
$(m_1 \neq \pm j_1~ {\rm and} ~m_2 
\neq \pm j_2)$.

Combining (\ref{ABBE2a}) and (\ref{PrimContr}) we can read off
\bea
A^{\tilde B}_{BB}(t) 
\approx \sum_{j,m_1=\pm m_2} \tilde n~ q^{2j^2}~
e^{-2 \tlam (|m_1|+|m_2|)}  e^{2(m_2 + m_1)t} ~, \qquad (q=q_1 q_2)
\label{ABBE2}
\eea
where we have defined 
\be
\tlam \equiv |\log(\sinlam)|
\ee
and absorbed all the numerical and phases in $\tilde n$. Different
from the series in (\ref{xSeries}) (with $x$ replaced by $-it$), which
is divergent and thus needs to be redefined for $t>\tau_\lambda$, 
we notice that in (\ref{ABBE2}) the summation over $j$ and $m$ is
convergent due to the factors of $q^{2j^2}$. 

Since the full-brane is time reflection symmetric, we will only focus
on the decay side $t>0$.
For $t\ll \tlam$~, (\ref{ABBE1}) is bigger than (\ref{ABBE2}) and we have
$A_{BB}(t) \approx 1$ because $f(t)$ has very weak $t$-dependence in
this region. This contribution starts to decrease around $t\sim \tlam$
and is gradually taken over by (\ref{ABBE2}). 
The leading term for
$t-\tau_\lambda \gg 2\pi s$ is especially simple. This comes from the
term with
$m_2=m_1=j-1\approx \frac{t-\tau_\lambda}{2\pi s}$ in (\ref{ABBE2}).
Up to numerical factors and phases we have\footnote{This is
reminiscent of the Euclidean cylinder, where we have a factor
of $e^{-y^2/4\pi s}$ if the string is stretched by a distance $y$. The
$t$-dependence here is like stretching a string in the time direction.}
\bea
A_{BB}(t) \sim e^{(t-\tau_\lambda)^2/\pi s} \qquad 
(t-\tau_\lambda \gg 2\pi s) ~.
\label{ABBtAppr}
\eea
Therefore $A_{BB}(t)$ in (\ref{ABBEa}) is well defined
for each $t$, but it grows rapidly as $t\rightarrow \infty$.
This is the quantum property of the tachyon matter if the
decaying 
brane does not completely go to the closed strings in the classical
level. Different contour of the $t$-integration in (\ref{ABBEa}) may
correspond to different choice of vacua\cite{Lambert:2003zr}. Since
all the contour will go through the large value of the real $t$, we
expect this qualitative behavior to be independent of the vacuum
choice. Physically closed strings will be created from such a one-loop
diagram and this rapid growth has to stop in a
time scale $\tau$ during which all the brane energy is emitted. 

Note that we have only considered the case where $s_{1,2}$ are
large. In principle we should integrate over the moduli $s_{1,2}$. We
see from (\ref{ABBtAppr}) that when $s$ decreases, $A_{BB}(t)$
increases, so it will only make the amplitude bigger. But the effect
we are unable to take into account in this paper is that when
$s_{1,2} \rightarrow 0$ the oscillator modes in $e^{iEX^0}$ are no
longer 
negligible. It is not clear, although unlikely, if the contribution of
these modes can 
exactly cancel the leading order in (\ref{ABBtAppr}).
Another complication is when $s\rightarrow 0$, we will encounter the
infrared divergence as suggested by (\ref{ABBtAppr}).
However if a time scale $\tau$
has to be imposed to this rolling tachyon evolution so that it is a
finite time process, we should have a natural infrared cutoff on $s$,
which is $s> 1/\tau$ (as well as infrared cutoff on closed string
propagation, so $s<\tau$).

\subsection{Closed string emission}
\label{SectClosed}
The time-dependent  boundary state provides a source for closed
strings. At the classical level, this is described by the one-point
function on the disk with the vertex operator representing the
physical
closed string state. The total energy emitted is proportional to the
sum of the square of the amplitudes associated with all the physical
states\cite{Itzykson:rh,Lambert:2003zr}. Because 
the contribution from the spacelike components are 
all equal to one for the
disk case\cite{Lambert:2003zr}, the amplitude-square for different
closed string states are
the same. They scale as $e^{-2\pi E}$ for large $E$. This exponential
damping factor is exactly cancelled by the exponential growing term in
the Hagedorn density $\sim
e^{2\pi E}$ and results in a power-law dependence of the total
emitted energy on the string level $n$.

Here we regard the one-point function on the cylinder as the quantum
correction to the above classical result. To proceed, we use the
simplest cutoff $\tau$ on the evolution of the rolling
tachyon. Therefore in (\ref{ABBE1a}) and (\ref{PrimContr}) we
integrate from $t=0$ to
$t=\tau$. The leading term for (\ref{ABBEa}) scales as $1/E$ for high
level states, which damps
much slower than the disk case.\footnote{For $\tau -\tau_\lambda \gg
2\pi s$, there is also a factor of $e^{(\tau-\tau_\lambda)^2/\pi s}$
coming from the leading term in (\ref{PrimContr}).} Imposing smoother
cutoff may reduce this 
amplitude. But as we will roughly see in Appendix \ref{AppClose}
that, the spatial parts of the amplitudes for different emitted closed
string 
states is no longer the same as they are in the disk case, they grow
rapidly as the string level increases. Combining with the
growth of the Hagedorn density, this will result in a rapid divergence
in the emitted total energy as the closed string level increases. This
may overwhelm the classical result
and reduce the mass level of the emitted closed strings if the string
coupling $g$ is not
sufficiently small. 

Clearly, to study more details, we need to have a better understanding
of the back reaction and more complete calculation of the loop
diagram. Other channels, such as the two-point function on the disk and
higher order diagrams also deserve further study.

\subsection{Adding a spectator brane}
In this subsection we consider the time evolution of a cylinder
diagram with one boundary given by Sen's rolling tachyon state and
another satisfying the Neumann boundary condition. As in the previous
discussion we will concentrate on the timelike part, the spacelike
part of the cylinder diagram is as
usual\cite{DiVecchia:1999rh}. Applied to the
superstring theory, this corresponds
to a one-loop diagram connecting a decaying brane and a stable
brane. On the stable brane, open strings will be created if string
coupling is considered. Since stable D-brane does not exist in the
bosonic string theory that we will study, one
can image that one brane is decaying much slower than another.

We have the choices of putting the vertex operator in the middle of
the cylinder or on
the Neumann boundary. We will calculate the latter as an example. We
again restrict to the long cylinder case.

The amplitude is 
\bea
\tilde A_{BN}(E) 
&\equiv& \int dt~e^{iEt} A_{BN}(t)
\nonumber \\
&\approx& \langle B| q^{L_0+\tilde L_0} e^{iEt} |N\rangle ~.
\eea
The Neumann boundary state is given by\cite{Callan:1994ub}
\bea
|N\rangle = \sum_j |j;0,0\rangle\rangle ~.
\eea

As in Sec.~\ref{SectEvolution}, the constraint 
\bea
j_1^2 - m_1^2 = j_2^2
\label{jmConstr}
\eea
($m_2=0$ in this case) is satisfied by the following several cases. 
For $m_1=\pm j_1$ and $j_2=0$, similar argument as in
Sec.~\ref{SectEvolution} leads to a contribution to $A_{BN}(t)$ as
\bea
q^{\half\frac{\partial^2}{\partial t^2}} f(t) ~.
\label{ABNt1}
\eea
This gives a relatively flat time evolution for $t<\tau_\lambda$,
which means that the tachyon has not started to roll quickly and the
boundary state is still close to Neumann boundary.
The $j_1=j_2=j$ and $m_1=m_2=0$ case gives
a time-independent term $\approx q^{\half}$ in
$A_{BN}(t)$, which is quite different from
(\ref{ABBtAppr}). Therefore the special case which we did not
consider in 
Sec.~\ref{SectRolling} becomes important here. These contribution will
become dominant for large $t$. It
is not difficult to see that the evolution for large $t$ is much
slower than (\ref{ABBtAppr}). For example there are
special values satisfying (\ref{jmConstr}) with $m_1 \neq 0$ and
$j_2 \neq 0$. Even if we assume that $j_1$ and $m_1$
is continuous, we will get $e^{(t-\tau_\lambda)^2/4\pi s}$ (at
$j_1\approx m_1 \approx (t-\tau_\lambda)/4 \pi s$) which is
much slower than (\ref{ABBtAppr}) due to the time-independence
of the Neumann boundary 
state. The situation is similar for cases with $j_1^2-m_1^2 \neq
j_2^2$.
In addition if these two branes are separated by a distance $y$ in the
transverse direction, there will be another suppression factor
$e^{-y^2/4\pi s}$ coming from the spacelike part of the diagram. 

Including the spatial and ghost parts of the open string vertex
operator, this 
one-point function on the Neumann boundary corresponds to the physical
open string emission on the spectator brane. We expect that a coherent
state of massive open strings will be created. At the same time, the
closed strings will also be created by inserting
the closed string vertex operator in the bulk of this cylinder. It
will be interesting to study these in more details and see how the
energy of the decaying brane is distributed between the open and
closed
strings. This will have applications in various brane inflation
models (see\cite{Quevedo:2002xw} for a review).

\section{Conclusion}
\label{SectCon}
In this paper we have studied the one-loop evolution in the rolling
tachyon background and the closed string emission from such a
diagram. We calculated the long cylinder case where we approximate
the vertex operator by its zero mode. In this case we show that the
one-loop diagram will grow rapidly in time. This indicates that, if
the 
tachyon matter survives the classical closed string emission, it will
be converted to closed strings at quantum levels. The short cylinder,
back reaction and other string emission channels have to be understood
better to study more details of such a process.

It should be straightforward to extend the analyses in this paper to
the half-brane
case\cite{Strominger:2002pc,Larsen:2002wc,Gutperle:2003xf} and the
superstring case\cite{Sen:2002in,Larsen:2002wc}. It will also be
interesting 
to study this when other backgrounds are present, such as the
electric
field\cite{Mukhopadhyay:2002en,Rey:2003xs,Rey:2003zj,Sen:2003xs,Nagami:2003yz}
or spacelike linear dilation\cite{Karczmarek:2003xm}.

\acknowledgments 
I would like to thank Brandon Bates, Pei-Ming Ho, Zongan Qiu, Gary
Shiu and Erick Weinberg for useful conversations. I am especially
grateful to Dan Kabat, Ashoke Sen and Charles Thorn for many helpful
discussions. This work was supported in part by the Department of
Energy under Grant No.~DE-FG02-97ER-41029.

\appendix
\section{One-point functions on an Euclidean cylinder}
\label{AppOnePoint}
In this appendix, we compute the one-point functions on an Euclidean
cylinder with
the Dirichlet or Neumann boundary condition and compare these with the
approximation we use in the paper. 

\subsection{DD boundary conditions}

\begin{figure}[htb]
\begin{center}
\epsfig{file=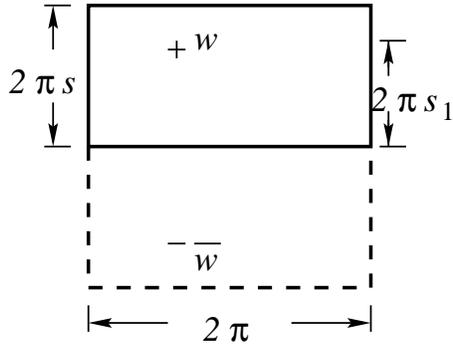, width=6cm}
\end{center} 
\medskip
\caption{The solid lines represents a cylinder of length $2\pi s$ and
the circumference $2\pi$. The left and right vertical boundaries are
to be identified. Here both horizontal boundaries satisfy the
Dirichlet boundary condition. The vertex operator is inserted at
$w$. The extended diagram including the dash lines and the image
of the vertex operator is the equivalent torus. The ``$+$'' and
``$-$'' signs represent the charges of the sources.}
\label{CylinderDD}
\end{figure}

We start with the case of a single D-brane with a
closed string tachyon vertex operator $e^{ikX}$. The length of the
cylinder is
$2\pi s$ and the circumference is $2\pi$. This corresponds to a strip
in Fig.~\ref{CylinderDD} with vertical boundaries identified. The
vertex operator is at $w=2i \pi 
s_1$. Because of the Dirichlet
boundary condition, this source have images of opposite charges
reflected by both horizontal boundaries. This is equivalent to a torus
with periodicities $2\pi$ and $4\pi s$ and two
opposite charges inserted at $w$ and $\bar w$ respectively.
The self-contraction Green's function is thus given
by\cite{Polchinski:rq}
\bea
G'_r(w,w) &=& \ln \left| 2\pi 
\frac{{\cal \theta}_1(\frac{w-\bar w}{2\pi},2is)}
{ {\cal \theta}_1' (0,2is) } \right|
- \frac{\left[{\rm Im}(w-\bar w)\right]^2}{ 8\pi s}  \nonumber \\ 
&=& \ln \left| 2\pi \frac{{\cal \theta}_1(2is_1,2is)}
{ {\cal \theta}_1' (0,2is) } \right|
- 2\pi \frac{s_1^2}{s}
~.
\eea
The one-point function is then
\bea
\langle e^{ikX} \rangle = C_{DD} e^{-\half k^2 G'_r(w,w)} ~,
\label{Onepoint}
\eea
where $C_{DD}$ is the vacuum amplitude of the cylinder (neglecting 
an overall factor of $q^{-1/12}$)
\bea
C_{DD} = \frac{1}{4 \pi^2 \sqrt{s} h(q^2)} ~,
\eea
where
\bea
h(q^2) = \prod_{n=1}^\infty \left( 1-q^{2n} \right) ~.
\eea

We now look at this amplitude in different limits. It is useful to
recall the formula\cite{Green:mn}
\bea
\frac{{\cal \theta}_1(\nu,\tau)} {{\cal \theta}_1'(0,\tau)} 
= \frac{i}{2\pi} \frac{1-z}{\sqrt{z}} ~\prod_{n=1}^\infty 
\frac{ (1-q^n z)(1-q^n z^{-1})} {(1-q^n)^2}  ~,
\label{ThetaForm}
\eea
with
\bea
q=e^{2i\pi\tau} ~, \qquad z=e^{2i\pi\nu} ~,
\eea
and it modular transformation
\bea
\frac{{\cal \theta}_1(\nu,\tau)} {{\cal \theta}_1'(0,\tau)} = 
-\tau e^{-i\pi\nu^2/\tau} 
\frac{{\cal \theta}_1(-\nu/\tau,-1/\tau)} 
{{\cal \theta}_1'(0,-1/\tau)}   ~.
\label{ThetaFormTran}
\eea

For $s_1,s_2 \gg 1$, using Eq.~(\ref{ThetaForm}), the one-point
function (\ref{Onepoint}) goes to 
\bea
\langle e^{ikX} \rangle &\rightarrow& C_{DD} 
e^{-\pi k^2 s_1 s_2/s }  ~.
\label{gg1Limit}
\eea
In this limit, we recover the amplitude (\ref{ADDk}) as we expected.

We next consider the cases where $s$ or $s_1$ gets small. In the first
case we consider $s_1 \ll 1$ and $s_1/s \ll 1$
for given $s$. This corresponds to bring the vertex operator close to
the boundary. In this case, 
\bea
\langle e^{ikX} \rangle \rightarrow C_{DD} (4\pi s_1)^{-k^2/2} ~.
\label{SmallLimit1}
\eea
If we evaluate (\ref{gg1Limit}) at this limit, we will get
$C_{DD}$. So the actual result (\ref{SmallLimit1}) differs from it by
a factor of $(4\pi s_1)^{-k^2/2}$, which diverges as $s_1 \rightarrow
0$.

The other case is that $s\ll 1$ but with $s_1$ comparable to $s$,
i.e.~$s_1/s \sim {\rm const}$. In this case the cylinder becomes a
long strip. This limit can be most easily studied
using the modular transformation (\ref{ThetaFormTran}) before using
(\ref{ThetaForm}). This gives
\bea
\langle e^{ikX} \rangle &\rightarrow& C_{DD} \left( 4s
\sin\pi\frac{s_1}{s} \right)^{-k^2/2} ~. 
\eea
This differs from (\ref{gg1Limit}) by a factor of
$\left( 4s\sin(\pi s_1/s) \right) ^{-k^2/2} \sim s^{-k^2/2}$.
The infrared behavior in these two cases are similar.

\subsection{DN boundary conditions}
\begin{figure}[htb]
\begin{center}
\epsfig{file=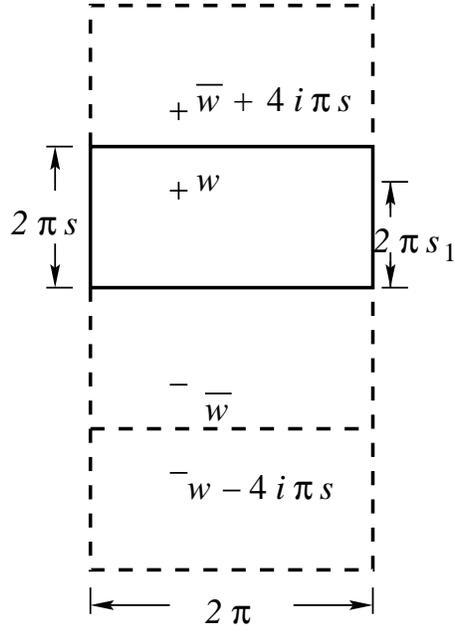, width=6cm}
\end{center} 
\medskip
\caption{Here the upper horizontal solid line represents Neumann
boundary, and the lower one represents Dirichlet boundary. They give
different image charges.}
\label{CylinderDN}
\end{figure}

We consider the cylinder with one Dirichlet boundary and one Neumann
boundary. We first consider the closed string insertion. This is
equivalent to have four sources at $w$, $\bar w$, ${\bar w} + 4i\pi s$
and $w-4i\pi s$ in a torus with periodicities $2\pi$ and $8\pi
s$ (Fig.~\ref{CylinderDN}). The mirror images reflected by the
Dirichlet boundary have the opposite charges, and those reflected by
the Neumann boundary have the identical charges.

The one-point function is given by
\bea
\langle e^{ikX} \rangle _{DN,c} =
C_{DN} ~ \left| 2\pi \frac{ {\cal \theta}_1(2is_1,4is) }
                          { {\cal \theta}_1'(0,4is) } 
  \frac{ {\cal \theta}_1(2is,4is) }
       { {\cal \theta}_1(-2is_2,4is) } \right|^{-k^2/2}
e^{\pi k^2 s_1} ~,
\label{OnepointDNc}
\eea
where $C_{DN}$ is the vacuum amplitude of the cylinder with DN
boundary condition,
\bea
C_{DN} = \frac{1}{2\pi} 
\prod_{n=1}^\infty \frac{1}{ (1+q^{2n}) } ~.
\eea

Taking the limit $s_1,s_2 \gg 1$, (\ref{OnepointDNc}) reduces to 
\bea
C_{DN} ~ e^{-\pi s_1 k^2}
\label{gg1LimitDNc}
\eea
This is the same as what we will get using the method in the paper
\bea
\langle D | q_1^{L_0 + {\tilde L}_0} ~e^{ikx}~ 
q_2^{L_0 + {\tilde L}_0} | N \rangle ~.
\eea

If the vertex operator gets close to the Dirichlet boundary
($s_1\rightarrow 0$, $s_1/s
\rightarrow 0$), or the cylinder becomes a long strip ($s\rightarrow
0, s_1/s \sim {\rm const}$), we have the similar infrared behavior
as in the last subsection with factors of $s_1^{-k^2/2}$ or
$s^{-k^2/2}$. But if the vertex gets close to the Neumann boundary
($s_2\rightarrow 0$, $s_2/s \rightarrow 0$), we will get a factor
which vanishes as $s_2^{k^2/2}$.

Last we consider the cylinder with a vertex operator at the
Neumann boundary. This is equivalent to a torus with opposite sources
at $2i \pi s$ and $-2i \pi s$. Since these sources are at the
boundaries, the intensity is doubled because of the overlapped image
sources. Similar calculation gives the one-point function
\bea
\langle e^{ikX} \rangle _{DN,o} = C_{DN}~
\left| 2\pi \frac{ {\cal \theta}_1(2is,4is) }
                 { {\cal \theta}_1'(0,4is) } 
\right|^{k^2} e^{\pi k^2 s} ~.
\label{OnepointDNo}
\eea
For large $s$, this gives
\bea 
C_{DN} ~ e^{-\pi k^2 s} ~,
\label{gg1LimitDNo}
\eea
which is the same as 
\bea
\langle D | q^{L_0 + {\tilde L}_0} ~e^{ikx}~ | N \rangle ~.
\eea
As $s \rightarrow 0$, (\ref{OnepointDNo}) goes to 
\bea
C_{DN} ~ (8s)^{-k^2} ~.
\eea
This is the similar infrared behavior we encountered previously.

\section{Closed string emission from one-loop}
\label{AppClose}
As mentioned in Sec.~\ref{SectClosed}, in order to get the total
emitted energy, we need to calculate the
sum of the square of the amplitudes of all possible close string
one-point functions. In this appendix we consider the contribution of
the one-loop diagram. We shall give a rough estimate of the lower
bound of this quantity and see its divergence.

For physical closed string states, the oscillator
modes of the time-like component can be gauged
away\cite{Lambert:2003zr,Hwang:1991an,Evans:1998qu}. The one-point
function then factorizes as
\bea
A = \langle e^{iEX^0} \rangle_{X^0}
\langle V_{sp} \rangle_{\vec x} ~.
\eea
The first factor is the  
timelike component that we have studied in
Sec.~\ref{SectEvolution}. The second factor is the spacelike
component with the closed string vertex operator $V_{sp}$. The ghost
part gives a
time and state-independent factor, which is not included here.
The evaluation of this part is more complicated than the disk
case. For example, the amplitude for different closed
string state is generally different, and the left and right
oscillators do not have to be identical. For simplicity, we will study
a subset of all possible closed string emission.

We consider the closed strings with oscillator modes only in one
spatial direction, for example considering the Dirichlet boundary
condition, and with zero transverse momenta and left-right identical
oscillators
\bea
V_{sp} = \prod_m \left( -\frac{2m}{m!^2} 
\partial^m X \bar\partial^m X \right)^{n_m} ~. 
\eea
For these strings, the
zero-modes contribution to the one-loop amplitude
\bea
\langle D | q_1^{L_0 + \tilde L_0} ~V_{sp}~ q_2^{L_0 + \tilde L_0}
|D\rangle 
\label{SpOneLoop}
\eea
is proportional to
\bea
\int_{-\infty}^{\infty} dp ~q^{p^2/2} = \frac{1}{\sqrt{s}} ~.
\label{SpZeroMode}
\eea
Let us only look at the lowest oscillator-modes in the corresponding
vertex operators. These modes are
\bea
V_{sp}  \rightarrow
\prod_m \left(m \alpha_1 \tilde \alpha_1 \right)^{n_m} ~,
\eea
evaluated at the world sheet coordinate $z=1$. 
To lowest order
in $q$, contribution of these modes
to the one-loop amplitude (\ref{SpOneLoop}) is given by
\bea
q_2^{2n} n! \prod_m m^{n_m} ~,
\label{SpOsciMode}
\eea
where $n\equiv \sum_m n_m$.
After including (\ref{SpZeroMode}) (\ref{SpOsciMode}) and
using the result discussed in Sec.~\ref{SectClosed} $\langle e^{ikX^0}
\rangle_{X^0} \sim \frac{1}{E} e^{(\tau-\tau_\lambda)^2/\pi s}$ 
for heavy
strings with $E = \sqrt{4N} ~(N\equiv \sum_m mn_m)$ and then 
squaring the amplitude,\footnote{The quantity we consider here is of
order $g^2$. There
are also cross terms of order $g$ between the one-loop and disk
amplitude. For massive strings these terms become smaller due to the
exponential fall-off of the disk 
amplitude as a function of $E$. Therefore for our purpose (to
see the divergent behavior of the heavy strings emission) we do not
consider them here.} we get
\bea
|A|^2 \propto \frac{1}{s} ~e^{2(\tau-\tau_\lambda)^2/\pi s}~ \frac{1}{N}~
\left( q_2^{2n} n! \prod_m m^{n_m} \right)^2 ~.
\label{SqAmp}
\eea
Actually before we square the amplitude, we should integrate over the
moduli $s_{1,2}$. Here to make a rough estimate we only integrated
over a unit length of
$s_{1,2}$ taking any value of $s_{1,2}$ as long as they are
large. Since the amplitude does not change sign as we change $s_{1,2}$
when $s_{1,2}$ are large, the integration over other region will only make
the amplitude bigger. But the effect of the small $s_{1,2}$ region is
unclear and is not taken into account in this paper.

In (\ref{SqAmp}) the factors in the bracket grow rapidly as the
level of the string states $N$ increases. For the same level we also have
the degeneracy of states characterized by the Hagedorn density. So
after summing over all possible
$\{ n_m \}$ for (\ref{SqAmp}), the total energy emitted will diverge
very rapidly as $N$ increases.

\end{document}